\documentstyle[epsfig,preprint,aps]{revtex}           
\begin{document}       
\tighten
\def\bea{\begin{eqnarray}}          
\def\eea{\end{eqnarray}}          
\def\beas{\begin{eqnarray*}}          
\def\eeas{\end{eqnarray*}}          
\def\nn{\nonumber}          
\def\ni{\noindent}          
\def\G{\Gamma} 
\def\D{\Delta}         
\def\d{\delta}          
\def\l{\lambda}          
\def\g{\gamma}          
\def\m{\mu}          
\def\n{\nu}          
\def\s{\sigma}          
\def\tt{\theta}          
\def\b{\beta}          
\def\a{\alpha}          
\def\f{\phi}          
\def\fh{\hat{\phi}}          
\def\y{\psi}          
\def\z{\zeta}          
\def\p{\pi}          
\def\e{\epsilon}           
\def\ve{\varepsilon}
\def\cd{{\cal D}}          
\def\cl{{\cal L}}          
\def\cv{{\cal V}}          
\def\cz{{\cal Z}}          
\def\pl{\partial}          
\def\ov{\over}          
\def\~{\tilde}          
\def\rar{\rightarrow}          
\def\lar{\leftarrow}          
\def\lrar{\leftrightarrow}          
\def\rra{\longrightarrow}          
\def\lla{\longleftarrow}          
\def\8{\infty} 
\def\pls{\partial\!\!\!/}
\def\bs{b\!\!\!/}
\def\ps{p\!\!\!/}
\def\As{A\!\!\!/}
\def\yb{\bar{\y}}
\def\Ds{D\!\!\!\!/}    
\newcommand{\fr}{\frac}          
          
\title{Lorentz and CPT Violating Chern-Simons Term\\ 
in the Derivative Expansion
of QED\thanks
{This work is supported in part by funds provided by the U.S.
Department of Energy (D.O.E.) under cooperative
research agreement \#DF-FC02-94ER40818.}}
          
\author{J.~-M. Chung\footnote{Electronic address: chung@ctpa03.mit.edu}
~and Phillial Oh \footnote{Electronic address:~ploh@newton.skku.ac.kr;
Present address: Department of Physics
and  Basic Science Research Institute,
Sung Kyun Kwan University, Suwon 440-746, Korea}}
\address{Center for Theoretical Physics\\
Massachusetts Institute of Technology\\
Cambridge, Massachusetts 02139\\
{~}} 
       
\date{MIT-CTP-2809,~~~~ December 1998}                  
\maketitle              
\draft              
\begin{abstract}           
\indent           
~~We calculate by the method of dimensional regularization and 
derivative expansion  the one-loop effective action for 
a Dirac fermion with a Lorentz-violating and CPT-odd 
kinetic term in the background of a  gauge field. 
We show that this term induces a  Chern-Simons 
modification to Maxwell theory. Some related issues are also 
discussed.
\end{abstract}              
                   
      
Violation of Lorentz symmetry, 
if it exists, will have a significant 
impact in our understanding of Nature and its symmetries.
There have been some recent attempts \cite{cfj,cg,ck}
to analyze the possible consequences of 
actual breaking of Lorentz invariance. 
One \cite{cfj} of the  proposals was a  
Lorentz-CPT non-invariant modification of electromagnetism by 
a Chern-Simons term  \cite{dj}.  
More recently, a  Lorentz 
violating extension of the standard model was presented, 
and possible  consequences including  relation to 
QED were explored in detail \cite{ck}. 

In this Report, we study QED with a Lorenz-violating 
CPT-odd term in the fermion sector, and show that
this term induces a finite Chern-Simons modification to
Maxwell theory in the one loop effective action.   
Recall that   Carroll, Field, and Jackiw \cite{cfj}
first proposed the following Lorentz and CPT violating 
Maxwell-Chern-Simons theory
\bea
\cl=-{1\ov 4}F_{\m\n}F^{\m\n}-j_\m A^\m-
{1\ov 2}l_\m~ ^*F^{\m\n}A_\n\;,
\label{csl}
\eea 
and this model was  reexamined in the 
context of Lorentz violating extensions of the 
standard model in Ref.~\cite{ck}.
In the above Lagrangian, $l_\m$ is a
constant 4-vector that picks out a preferred direction 
in space-time, thereby 
violating Lorentz invariance. CPT symmetry is absent as well. 
Possible origin of such a Chern-Simons term is that 
it can be induced by 
radiative corrections \cite{rw}
from a Lorentz and CPT violating fermion sector.
With this in mind, let us consider the Dirac fermion 
propagating in Lorentz and CPT non-invariant manner 
in the background of a photon field;
\bea
\cl=\yb[i\pls-m-\g_5\bs-e\As]\y\;. \label{fl}
\eea
Here  $b_\m$ is a constant 4-vector, and 
$b_\m\yb\g^\m\g_5\y$ is the Lorentz violating, CPT-odd term. 

Such a term was considered as a possible extension 
to the standard model
by Colladay and Kosteleck\'{y} \cite{ck}.
They also  calculated by the diagrammatic method the 
radiative corrections 
to vacuum polarization involving the $b_\m$ term. 
The  computation of vacuum polarization two-point diagram
with  an extra insertion of the factor 
$-ib_\m\g_5\g^\m$ on one internal
fermion line leads to  the triangular anomaly \cite{abj,ad} diagram
in which there is zero momentum
transfer to the axial-vector leg and the axial vector 
is replaced with a vacuum expectation value. 
The result is finite, but
there is an ambiguity due to  linear divergence of 
momentum integration.
However, a more careful analysis \cite{jk} shows that not only 
such  superficial
linear divergences cancel each other, but also a definite
value can be calculated for the vacuum polarization diagram 
involving the $b_\m$ term.  

Here, we directly compute the 
effective action of the model in Eq.~(\ref{fl}).
Using dimensional regularization \cite{tv}
and the derivative expansion method of 
Ref.~\cite{ccz}, 
we  show  that the 
Chern-Simons term of Eq.~(\ref{csl}) is induced 
in the effective action with a fixed coefficient. 
The  one loop effective action  $S_{\rm eff}$ is given by
\bea
S_{\rm eff}=\int d^4x\cl_{\rm eff}=
-i{\rm Tr}\ln [i\pls-m-\g_5\bs-e\As]\;.
\label{effective}
\eea
To carry out the trace calculation, we first use the trace identity;
\bea
{\rm Tr}\ln [i\pls-m-\g_5\bs-e\As]=
{\rm Tr}\ln [i\pls-m-\g_5\bs]-\int^1_0dz~
{\rm Tr}{1\ov i\pls-m-\g_5\bs-ze\As(x)}e\As(x)\;.
\label{loopi}
\eea
Note that $\pl_\mu$ and $A_\nu(x)$ do not commute, and 
to perform the  momentum space integration
of the second term in Eq.~(\ref{loopi}), we use the prescription of
Ref.~\cite{ccz} in the denominator of Eq.~(\ref{loopi});
\beas
i\pls\rightarrow \ps,~~~~~   
\As(x)\rightarrow \As(x-i{\pl\ov \pl p})\;.
\eeas
Using this, we rewrite $S_{\rm eff}$  as
\beas
S_{\rm eff}=-i{\rm Tr}\ln \Ds_0+i\int_0^1 dz
\int d^4x\int{d^4p\ov (2\p)^4}
{\rm tr}\biggl({1\ov \ps-m-\g_5\bs-
ze\As(x-i{\pl\ov \pl p})}e \As(x)\biggr)\;,
\eeas
where
\beas
\Ds_0=i\pls-m-\g_5\bs\;.
\eeas
$\Ds_0$ propagator has been analyzed in detail in 
the second paper of 
Ref.~\cite{ck}. We study the second term of $S_{\rm eff}$, 
and look for first order  derivative terms which are 
linear  in $\bs$ and 
quadratic in  $A_\m$'s. Using the operator expansion 
\beas
{1\ov A-B}=A^{-1}+A^{-1}BA^{-1}+\cdots\;,
\eeas
we can  extract these to be
\bea
\int d^4x\cl_{\rm eff}^{(b)}\equiv 
-i{e^2 \ov 2}\int d^4x\int {d^4 p\ov (2\p)^4}{\rm tr}
\biggl[
{1\ov \ps-m }i\pl_\m \As{\pl\ov \pl p_\m}{ 1\ov \ps-m}\g_5
\bs{1\ov\ps-m}\As\nn\\
+{ 1\ov \ps-m}\g_5\bs{1\ov\ps-m}
i\pl_\m\As{\pl\ov \pl p_\m}{1\ov\ps-m}
\As\biggr]\;.\label{el} 
\eea  
Using 
\beas 
{\pl\ov \pl p_\m}{1\ov \ps-m}=-{1\ov \ps-m}\g^\m{1\ov \ps-m}\;,
\eeas
we first rewrite Eq.~(\ref{el}) into
\bea
\int d^4x\cl_{\rm eff}^{(b)}&=& 
-{e^2 \ov 2}\int d^4x\int {d^4 p\ov (2\p)^4}{\rm tr}
\biggl[\g_5\bs
{1\ov \ps-m }\biggl(\As {1\ov \ps-m}\pl_\m 
\As{ 1\ov \ps-m}\g^\m{1\ov \ps-m}\nn\\
&+&\g^\m{1\ov \ps-m}\As{1\ov \ps-m}\pl_\m \As{1\ov\ps-m}
+\pl_\m\As{1\ov\ps-m}\g^\m{1\ov \ps-m}\As{1\ov \ps-m}
\biggr)\biggr]\;.\label{drel} 
\eea  

The momentum integral has terms which diverge logarithmically, and
in order to regularize the expression, we use the dimensional 
regularization \cite{tv}. We use the rule \cite{ra}
that we first perform the loop integrals in arbitrary $d$ dimensions, and 
identify the potential divergences as $d\rightarrow 4$. Then, 
we are left with traces of
$\gamma$-matrices containing $\gamma_5$.
We will take these traces directly in 4 dimensions. 
The traces are basically of three types,
\bea
 {\rm tr}(\gamma_5 \gamma_\a \gamma_\b \gamma_\g \gamma_\d)\;,~~
 {\rm tr}(\gamma_5 \gamma^\a \gamma_\b \gamma_\a \gamma_\d 
\gamma_\mu\gamma_\nu)\;,~~
{\rm tr}(\gamma_5 \gamma^\a \gamma_\b \gamma_\g \gamma_\d
\gamma_\a\gamma_\nu)\;.\label{gim}
\eea
The second and third terms contain $\gamma$ matrices contracted, and
we eliminate such contractions by using the identities
\beas
\gamma^\a \gamma_\b \gamma_\a=-2\gamma_\beta,~~~
\gamma^\a \gamma_\b \gamma_\g \gamma_\d
\gamma_\a=-2\gamma_d\gamma_\g\gamma_\b.
\eeas
Then, all the trace 
computations involving $\gamma_5$
reduce to terms of the first type in Eq.~(\ref{gim}), which is
equal to $-4i\epsilon_{\alpha\beta\gamma\delta}$. 
Under this regularization scheme, potential divergences in the momentum 
integration come from the following type of integral;
\bea
\int{d^dp\ov (2\p)^d}{
p_\a p_\b p_\g p_\d\ov (p^2-m^2)^4}
&=&{i(g_{\a\b}g_{\g\d}+g_{\a\g}g_{\b\d}+g_{\a\d}g_{\b\g})
\ov (4\p)^{d/2}}
{\G(2-d/2)\ov 24}\biggl({1\ov m^2}\biggr)^{2-d/2}\;.\label{di2}
\eea
But the leading logarithmic divergences cancel in each of the three 
terms of Eq.~(\ref{drel}) separately, and we obtain a 
completely finite result.
The remaining integrals converge like $1/p^2$ 
as $p\rar \8$ and $d\rar 4$. 
By using the following momentum integration in Minkowski space,
\beas
\int{d^4p\ov (2\p)^4}{1\ov (p^2-m^2)^4}&=&{i\ov 96\p^2}{1\ov m^4}\;,\nn\\
\int{d^4p\ov (2\p)^4}{p_\a p_\b\ov (p^2-m^2)^4}
&=&-{ig_{\a\b}\ov 192\p^2}{1\ov m^2}\;,
\eeas
we obtain the following effective Lagrangian
\bea
\cl_{\rm eff}^{(b)}={3e^2\ov 16\p^2}b_\m~^*F^{\m\n}A_\n\;.
\label{bfa}
\eea

Comparing this effective Lagrangian with Eq.~(\ref{csl}), 
we find that the Lorentz violating
CPT odd term $\yb\g_5\bs\y$ produces  the Chern-Simons term with 
coefficient
\bea
l_\m=-{3e^2\ov 8\p^2}b_\m\;.
\label{mass}
\eea
Note that the momentum integral in our  derivative expansion of 
Eq. (\ref{el}) has  leading logarithmic divergences,
but the result is finite 
and it is consistent with the computation of the vacuum polarization
diagram with the fermion propagator $S_{\rm F}=
i(\ps-m-\gamma_5\bs)^{-1}$
\cite{jk}. Also, Eq.~(\ref{bfa}) is 
independent of the mass term and the 
result is  highly reminiscent of the finite anomaly contribution to 
the divergence of axial vector current \cite{abj,ad}. 

If we include all the fermion species of the standard model, with each
having electromagnetic coupling strength $q_f^2$, 
 the induced Chern-Simons coefficient becomes
\bea
l_\m=-{3\ov 8\p^2}\sum_{f}q_f^2b^f_\m\;,
\label{grand}
\eea
where the sum over $f$ extends over all the leptons and quarks.
If the coupling coefficient $b^f_\mu$ is indeed generated
as the vacuum expectation values of an 
effective axial vector $<A^a_{\mu5}>$
which multiplies the axial-vector bilinears of fermions, 
as was suggested in 
Ref. \cite{ck}, we would have
$b_\mu^f=\sum_a g^a_f<A^a_{\mu 5}>$,  
where the index $a$ ranges over the species of fermion which have
such axial vector coupling, and $g^a_f$ is the associated coupling. 
Then the above equation for $l_\m$ is equivalent to
$\sum_{f}q_f^2g^a_f$,
which must vanish according to the anomaly cancellation 
condition \cite{ck}.  
However, it is also possible that
the Lorentz-violating term involving $b^f_\mu$ may not 
couple to fermions  through the vacuum expectation value of 
an axial vector. 

In conclusion, we found that the derivative expansion method
for effective action with the dimensional regularization
yields  a finite Chern-Simons 
modification to Maxwell theory, if the standard
Lagrangian is augmented by a  Lorentz-violating 
CPT-odd term in the fermion sector.
We used the dimensional regularization method, and  
the coefficient of the induced
Chern-Simons term is fixed and non-vanishing in this scheme.
But in the Pauli-Villars regularization,
the coefficient is zero, because  Eq. (\ref{bfa})
is mass independent \cite{ck,jk}. 
Hence the result depends on the regularization scheme one
chooses to adopt. 
Chern-Simons modification of Maxwell theory predicts 
a preferred direction of space-time and birefringence of  
photon propagation, which was compared with experiment \cite{cfj}.
The currently available experimental data \cite{noevid} rules out
the occurrence of this  Lorentz violating term in Nature.  
But this does not necessarily mean that $b_\mu^f$ 
vanishes separately for each $f$ in Eq. (\ref{grand}).
The details of the fermion sector will be determined by the 
ultimate theory.\\

~~~~~~~~~~~~~~~~~~~~~~~~~~~~~~~~~~~~~~~~~~~~~NOTE ADDED
\\

1. Shortly after the completion of this work, a no-go theorem 
by  Coleman and Glashow  which states
that $l_\mu$ vanishes to  first order in $b_\mu$ for any gauge
invariant CPT-odd interaction appeared \cite{cg2}. 
The issue of evading the no-go theorem  for the induced Chern-Simons
term has been subsequently discussed by 
Jackiw and Kosteleck\'{y} in Ref. \cite{jk}.
They calculate nonperturbatively the vacuum polarization tensor.
Upon differentiation of their formula with respect to the
external momentum which is then set zero, one arrives at our 
expression (\ref{drel}) and our final answer agrees with theirs. 
 
2.  There appeared several other computations on the subject
\cite {seve} indicating that the induced coefficient is finite but
undetermined perturbatively, and actual  determination of the
coefficient needs experimental input \cite{raja}.
We briefly mention connection with axion physics in this context.
Note that (\ref{bfa}) can be regarded as 
an effective interaction $\sim  a F_{\mu\nu} 
\tilde{F}^{\mu\nu}$  between photon and infinitely thick axion 
domain wall $a(x)= b_\mu x^\mu$ which has
uniform space derivative. This axion-photon interaction can be 
induced by  derivative axion-fermions interaction of the form $\sim
\partial_\mu a \bar\psi\gamma^\mu \gamma_5\psi$
through anomaly \cite{yao}, 
and the induced coefficient depends on the details of 
the particular  model employed and its experimental result \cite{axion}. 
However, since current experiment \cite{noevid} 
rules out the  induced Chern-Simons coefficient,
such infinite thick domain wall configuration is unrealistic.
We  acknowledge the referee for raising this issue.

\acknowledgements
We would like to thank Prof. R. Jackiw for suggesting this problem,
useful discussions, and pointing out an error in our earlier
evaluation of the integral.
We also thank Prof. V. A. Kosteleck\'{y}
and K. Choi for useful correspondences,
and Prof. S. Coleman for  discussion on their  no-go theorem.
JMC is supported in part by the Postdoctoral Fellowship of KOSEF.
~PO is supported by the Korea Research Foundation
through program 1998-015-D00034.

\end{document}